\documentclass[12pt]{article}

\usepackage{newtxtext,newtxmath}
\usepackage{graphicx}
\usepackage{mathtools}
\usepackage[hidelinks]{hyperref}  % clickable cross-referencing
\usepackage{bm}         % bold greek letters
\usepackage[letterpaper,margin=1in]{geometry}
\usepackage{enumitem}
\usepackage[svgnames]{xcolor}
\setlist[enumerate,itemize]{itemsep=0pt}

\renewenvironment{abstract}
	{\quotation}
	{\endquotation}

\date{}

\makeatletter
\renewcommand{\fnum@figure}{\textbf{Figure \thefigure}}
\renewcommand{\fnum@table}{\textbf{Table \thetable}}
\makeatother

\usepackage{natbib}
\usepackage{url}

\newcommand{\T}{^{\mathsf{T}}}
\newcommand{\B}[1]{\if#1\relax\bm{#1}\else\mathbf{#1}\fi} % bold text

\def\scititle{
	Controlling Complex Systems
}
\title{\bfseries \boldmath \scititle}

\author{
	Marco Coraggio$^{1}$,\and
	Davide Salzano$^{1}$\and Mario di Bernardo$^{1,2,\ast}$\and
    \\
    \small$^{1}$Scuola Superiore Meridionale, Naples, 80134, Italy
    \\
    \small$^{2}$Department of Electrical Engineering and Information Technology,\\ \small University of Naples Federico II, Naples, 80128, Italy. 
    \and
	\small$^\ast$Corresponding author. Email: mario.dibernardo@unina.it\and
}

\begin{document} 

% Insert the title and author list
\maketitle

% Abstract, in bold
% There are strict length limits, and not all formats have abstracts.
% Consult the journal instructions to authors for details.
% Do not cite any references in the abstract.
\begin{abstract}\bfseries\boldmath%
\noindent This chapter provides a comprehensive overview of controlling collective behavior in complex systems comprising large ensembles of interacting dynamical agents. Building upon traditional control theory's foundation in individual systems, we introduce tools designed to address the unique challenges of coordinating networks that exhibit emergent phenomena, including consensus, synchronization, and pattern formation. 
We analyze how local agent interactions generate macroscopic behaviors and investigate the fundamental role of network topology in determining system dynamics.
Inspired by natural systems, we emphasize control strategies that achieve global coordination through localized interventions while considering practical implementation challenges. The chapter concludes by presenting novel frameworks for managing very large agent ensembles and leveraging interacting networks for control purposes.
\end{abstract}

\bigskip


\medskip

\noindent\textbf{Key points}:
\begin{itemize}[noitemsep,topsep=0pt]
    \item The most common formalism to model complex systems is that of a graph---where vertices represent agents and edges represent the interaction between them---with agent dynamics modeled via ordinary differential equations.
    \item Controlling complex systems typically means enforcing a specific collective behavior, ranging from consensus and synchronization to pattern formation and herding.
    \item Control strategies for complex systems can be either centralized or distributed, and can often be categorized as either node control, edge control, or structural control.
    % \item Recent control strategies can transcend this traditional categorization, integrating elements from multiple paradigms. 
    % Examples include continuification-based control and shepherding strategies.
    \item Tools for establishing stability and convergence of control strategies include Lyapunov theory, contraction theory, and the master stability function approach.
\end{itemize}

\medskip

{\color{Maroon}
\noindent\textbf{Please, cite as}: M. Coraggio, D. Salzano, and M. di Bernardo, “Controlling complex systems,” in \emph{Encyclopedia of Systems and Control Engineering}, Springer, 2025, doi: 10.1016/B978-0-443-14081-5.00167-7.}

\tableofcontents

%-----------------------------------
\section{Introduction and Motivation}

Modern technological advances have fundamentally changed our perspective on feedback control. 
While classical control theory has primarily focused on regulating individual dynamical systems, contemporary challenges increasingly involve coordinating and controlling large groups of interconnected systems. 
From power distribution in smart grids to autonomous vehicle formations, from biological cell networks to social systems, we encounter scenarios where multiple agents must coordinate their behavior to achieve common objectives \citep{neba:06,st:01}.

Systems comprising multiple interacting units possess a distinctive characteristic that distinguishes them from traditional dynamical systems: they exhibit emergent collective behaviors that cannot be \emph{easily} derived from the analysis of individual components. These systems, known as \emph{complex systems}, represent a fundamental shift in our understanding of dynamics. This concept emerged in the 1940s \citep{weaver1948science}, and while its precise definition remains debated without complete consensus, scholars broadly agree that collective behavior arising from local interactions represents the fundamental characteristic of such systems \citep{vicsek2002bigger,ladyman2013complex,torres2021and,bianconi2023complex,estrada2024complex}. 
Moreover, the concept of interactions leading to emergent behavior is fundamentally linked to \emph{feedback} in control theory \citep{astrom2021feedback}.
Indeed, complex systems can be viewed as networks of numerous feedback loops, their \emph{degree of complexity} determined by both the number and intricacy of these interconnections.
These systems are inherently \emph{multi-scale}, as microscopic interactions between agents generate macroscopic collective behaviors observable at the population level.

The simplest example of such collective behavior is consensus, where all agents converge to a common equilibrium state \citep{OlFa:07}. 
A more general phenomenon, often observed in both natural and engineered systems, is synchronization, where an ensemble of interconnected nonlinear agents converges towards the same time-varying trajectory \citep{piro:01}. 
Beyond these fundamental phenomena, collective behavior manifests in diverse forms including pattern formation, platooning, and clustering.

The development of effective strategies for controlling collective behavior of complex multi-agent systems has become increasingly critical in addressing contemporary challenges.
These include the transformation of traditional power systems into smart grids and energy communities \citep{dobson2013,dorfler2013}, the engineering of resilient and sustainable supply chains \citep{hearnshaw2013complex}, the implementation of smart sensorized cities \citep{wang2017vehicular} and transportation networks \citep{alam2015heavy}, the advancement of synthetic biological control systems \citep{del2018future}, and the evolution of coordinated swarm robotic systems \citep{annaswamy2023control}.

The collective behavior emerging from large ensembles of interacting dynamical systems is strongly influenced by the structure of the interaction network among its constitutive agents \citep{liu2016}. 
These networks typically diverge from simple regular lattices \citep{ne:03, newman2018networks}, exhibiting intricate topological features that profoundly shape system behavior.
The interactions between agents further complicate the dynamics, as they can be time-varying, intermittent, subject to delays, or governed by nonlinear functions \citep{cortes2006finitetime,meng2018synchronization, almeida2017synchronization}.

Natural systems provide compelling examples of efficient control strategies.
Biological systems, in particular, demonstrate how global coordination can emerge through the influence of a remarkably small number of agents \citep{chen2014f}.
This principle of achieving global control through localized actions on select agents holds profound implications for technological applications. 
However, its implementation requires addressing fundamental questions: how to achieve and maintain desired collective behaviors using only local information, how to identify which agents to control and with what intensity, and how to formally demonstrate convergence to desired behaviors.
Building on these foundational challenges, many pressing applications, such as search and rescue operations \citep{queralta2020collaborative} or traffic control \citep{siri2021freeway}, require a group of controller agents to work together effectively in order to shape the behavior of a separate group of target agents. Recent research has focused on this idea of harnessing complex systems for control, emphasizing both the design of the controller populations and the specific conditions needed to successfully coordinate the target behaviors; a notable example is that of shepherding problems as discussed in \cite{lama2024shepherding}.

\begin{figure}[t]
    \centering
    \includegraphics[width=1\linewidth]{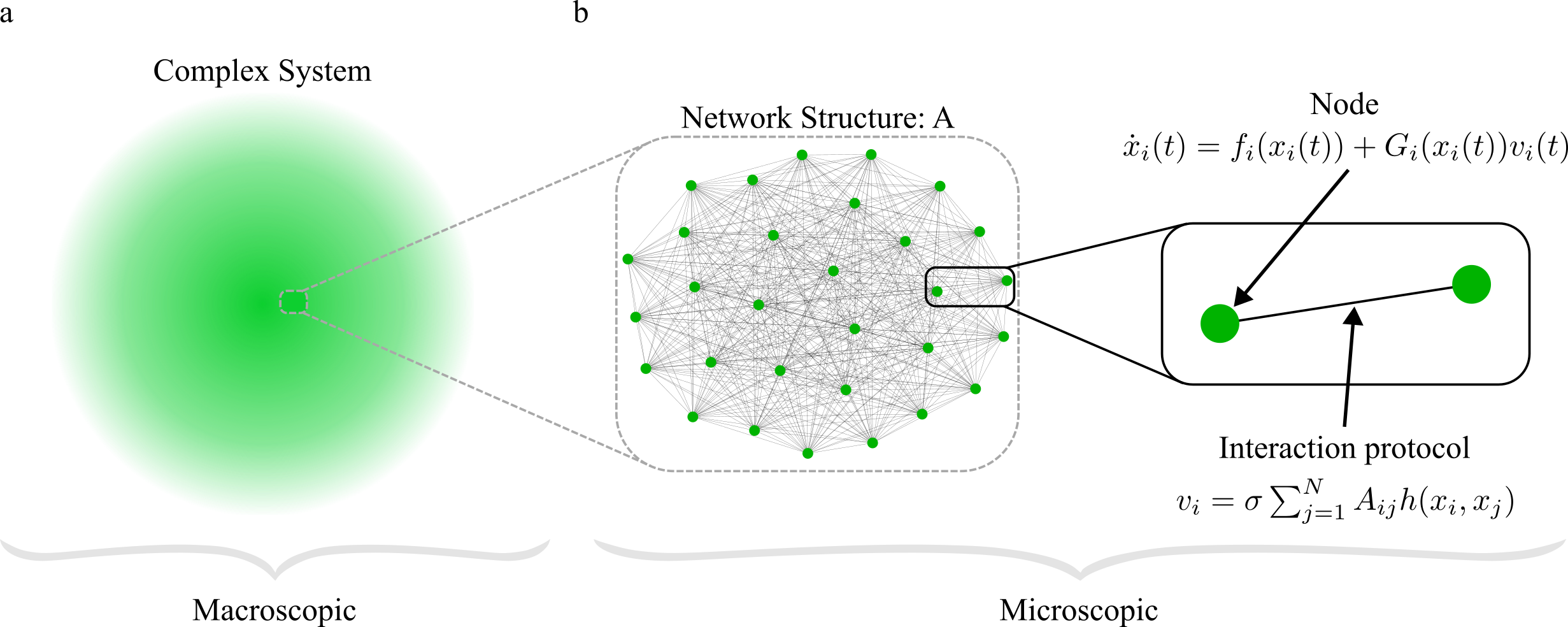}
    \caption{\textbf{Fundamental Elements of a Complex System.} \textbf{a.} Macroscopic description of a complex system. 
    Here, the identity of the agents composing the network is lost in favor of a description of the behavior that emerges at a system level.
    \textbf{b.} Microscopic description of a complex system. The key elements are the agents of the network and their interaction protocols (left panel), and the network describing the interconnections between agents (right panel).}
    \label{fig:Complex_System_description}
\end{figure}

Practical implementation must address several critical challenges that arise in real-world applications.
Heterogeneity among agents presents a fundamental difficulty, particularly in biological applications where identical components are rare \citep{del2018future}. 
This is compounded by communication constraints and the reality of only having access to partial information about the system state \citep{hespanha2007survey,zhang2020networked}.
Environmental disturbances and measurement noise further complicate the control problem, while the dynamic nature of many networks leads to time-varying topologies and interaction patterns that must be handled robustly. These practical considerations represent significant hurdles that must be overcome to translate theoretical control strategies into effective real-world solutions.

An emerging frontier is the development of strategies for systems comprising an extremely large number of agents. In such scenarios, the classical description of complex systems, which specifies the dynamics and interactions of each individual agent, becomes both analytically and computationally intractable. This has led to the development of new frameworks that address these systems directly at the population level through continuum descriptions \citep{nikitin2021continuation}.

This chapter presents a systematic exploration of the problem of controlling complex network systems. While the field is vast and rapidly evolving, we focus on control strategies that are general in their formulation and can be proved formally to converge. For broader perspectives on specific applications and alternative approaches, readers are referred to comprehensive reviews in both control theory and physics literature \citep{liu2016,Me:07,su2013,d2023controlling}.
The subsequent sections develop these concepts systematically, moving from basic modeling frameworks to advanced control strategies. We examine both theoretical foundations and practical implementations, with particular attention to emerging applications and future challenges in this exciting field.

%----------------------------------
\section{Modeling Complex Systems}

Understanding and controlling complex systems requires a systematic modeling framework for the latter.
A complex system can be understood as being the ensemble of \emph{units} (the atomic entities in the system) and the \emph{interactions} between them.
The units are often called \emph{systems}, \emph{nodes}, or \emph{agents}, while the the interactions are also referred to as \emph{communication}, \emph{(inter-)connections} and \emph{links}.
Mathematically, a complex system can be modeled through three fundamental components (see Figure \ref{fig:Complex_System_description}):
(i) the dynamics of the agents, 
(ii) the structure of their interconnections, and
(iii) their interaction protocols.
This description enables capturing both the microscopic behavior of individual agents and the emergence of macroscopic collective phenomena exhibited at a network level.
We expound each of the three components below.

%--------------------------
\subsection{Agent Dynamics}

In most control problems, it is assumed that each agent has a state that can change in time (see Figure \ref{fig:Complex_System_description}b, right panel).
Several mathematical formalisms are used to describe such dynamics, including (ordinary, delay, stochastic, and partial) differential equations, difference equations, Markov decision processes, and probability density/mass functions; yet, the most common formalism is arguably that of ordinary differential equations.
In that case, each agent is described as a nonlinear affine dynamical system, i.e., 
\begin{equation}\label{eq:general_network}
\dot{x}_i(t) = f_i(x_i(t)) + G_i(x_i(t)) v_i(t), \qquad i \in \{1,\ldots,N\},
\end{equation}
where,
$t \in \mathbb{R}$ is continuous time, 
$x(t) \in \mathbb{R}^n$ is the agent's state,
$f_i:\mathbb{R}^{n}\to\mathbb{R}^{n}$ is the \emph{agent}'s (also known as \emph{individual}, \emph{internal} or \emph{intrinsic}) \emph{dynamics},  
$v_i(t) \in \mathbb{R}^m$ is the input coming from the agent's  \emph{neighbors}---i.e., other agents with whom it is connected---$G_i:\mathbb{R}^n\to\mathbb{R}^{n \times m}$ describes how the agent responds to interaction,
and $N \in \mathbb{N}$ is the number of agents in the network.

The choice of the agent dynamics $f_i$ depends heavily on the application, with the complexity of the mathematical model varying according to the specific context.
In the simplest case, agents can be modeled as simple or higher-order integrators \citep{ren2008consensus}.
However, often more sophisticated models are required to capture the key dynamical features of the system. 
For example, autonomous vehicle applications generally involve second-order motion dynamics to capture both position and velocity states \citep{ye2018modeling}; in power systems, agents are represented by nonlinear oscillators describing power generators that must maintain synchronization \citep{hill2006power}; on the other hand, biological networks present arguably one of the most challenging scenarios, where agents represent complex nonlinear and stochastic cellular processes with intricate interaction patterns.
Examples of applications modeled via different frameworks include gene regulatory networks, often described using Markov chains or stochastic differential equations \citep{gillespie2000chemical}, 
systems of faults, modeled via discontinuous dynamical systems \citep{burridge1967model} and communication networks, typically represented through difference equations \citep{zhu2010discrete}.
The choice of appropriate agent dynamics modeling is context-dependent and must strike a good trade-off between model fidelity and analytical tractability.

%-----------------------------
\subsection{Network Structure}

The second essential component of a complex system is the \emph{network structure}, which describes the interconnections between agents.
Most commonly, interactions are assumed to be \emph{pair-wise}, i.e., each one involving exactly two agents.
In this case, the structure is  represented by a \emph{graph} \citep{godsil2013algebraic}, say $\mathcal{G} = (\mathcal{V}, \mathcal{E})$, where $\mathcal{V} = \{1, \dots, N\}$ is the set of \emph{vertices} (the agents), and $\mathcal{E} \subseteq \mathcal{V} \times \mathcal{V}$ is the set of \emph{edges} (the interconnections), see  Figure \ref{fig:Complex_System_description}b, left panel.
Here we assumed an \emph{undirected} graph (the presence of a connection from  agent $i$ to $j$ also implies a connection from $j$ to $i$), but \emph{directed} graphs are also common; in \emph{weighted} graphs---as opposed to the \emph{unweighted} ones assumed here---each edge also has a scalar value associated to it, typically representing a meaningful quantity in the application of interest (importance, capacity, travel time, etc.).

Graphs can be represented algebraically through two matrices.
The first one is the \emph{adjacency matrix} $A \in \{0, 1\}^{N \times N}$, with $A_{ij} = 1$ if an edge from $i$ to $j$ is present and $0$ otherwise.
Note that different authors can adopt different conventions on whether such edge means that agent $i$ influences $j$ or that $i$ is influenced by $j$.
The second matrix is the  \emph{graph Laplacian matrix} $L \in \mathbb{Z}^{N \times N}$, given by $L = D - A$, where $D \in \mathbb{N}^{N \times N}$ is the \emph{degree matrix}, which is diagonal and with $D_{ii}$ equaling the \emph{degree} of node $i$, i.e., the number of connections involving agent $i$ \citep{newman2018networks}. 
Network graphs can also be characterized macroscopically via a wide variety of descriptors, such as the \emph{average degree}, the \emph{clustering coefficient}, \emph{path lengths}, \emph{diameter}, different \emph{node centrality measures} and \emph{community structure}; for detailed descriptions, see \cite{bola:06,newman2018networks}.

Elementary well-known graphs typically used for model validation include \emph{paths}, \emph{rings}, \emph{stars}, \textit{wheels},
\emph{complete}, and \textit{$k$-nearest neighbors} \citep{godsil2013algebraic}.
On the other hand, it has been found that in applications certain kinds of structures tend to appear more consistently.
These include
\emph{random graphs} (built via the Erdős-Rényi model \citep{erdds1959random}), 
\emph{lattices}, that can be drawn as regular tilings,
\emph{small-world networks}, characterized by high clustering and short path lengths \citep{Wieland2011}, 
\emph{scale-free networks} exhibiting power-law degree distributions \citep{Watts:1998p307,liu2016}, and hierarchical or modular structures \citep{ravasz2002hierarchical}.

In the recent years, two more advanced modeling frameworks gained traction: multilayer networks, and networks with higher-order interactions.
In \emph{multilayer networks}, multiple graphs are used to describe the network; connections between agents in different graphs are possible, and if the vertices sets are the same for all graphs, the network is said to be \emph{multiplex} \citep{bianconi2018multilayer}. 
This formalism is used for example to describe multimodal transportation networks, where each mode of transportation (e.g., buses, subway, taxis) is modeled via a different graph, or social networks where each platform (e.g., Facebook, Instagram, X) is a graph, and more \citep{boccaletti2014structure}. 
Networks with interactions that involve more than two agents (\emph{higher order interactions}) are formalized via \emph{hypergraphs}.
These networks model, for instance, scientific authorship and biological processes \cite{boccaletti2023structure}.
Special cases of hypergraphs are \emph{simplicial complexes}, where the presence of a connection between a certain number of agents implies the presence of all possible connections between any proper subset of these agents.

%---------------------------------
\subsection{Interaction Protocols}

The third key element is the protocol through which agents exchange information and influence each other over existing connections (see Figure \ref{fig:Complex_System_description}b, right panel). 
Assuming pair-wise (unweighted) interactions, a common mathematical description for the interaction term $v_i$ in \eqref{eq:general_network} is
\begin{equation}\label{eq:coupling}
v_i = \sigma\sum_{j=1}^N A_{ij}h(x_i,x_j),
\end{equation}
where $\sigma \in \mathbb{R}$ represents the \emph{coupling} (or \emph{interaction}) \emph{strength} (or \emph{gain}), $A_{ij} \in \{0, 1\}$ is the $(i,j)$-th element of the adjacency matrix, and $h(x_i,x_j) : \mathbb{R}^{n} \times \mathbb{R}^{n} \to \mathbb{R}^m$ describes the \emph{interaction protocol}.

The interaction protocol is \emph{diffusive} if it can be written as $h(x_j - x_i)$.
In particular, a simple and common choice for $h$  is a \emph{linear diffusive coupling}, i.e., $h(x_i,x_j) = x_j - x_i$.
However, real systems often exhibit more complex interaction patterns that go well beyond simple linear protocols. 
Time delays inevitably arise in communication or actuation \citep{hespanha2007survey,zhang2020networked}, while coupling strengths between agents are rarely uniform, requiring heterogeneous gains ($\sigma_{ij}$ instead of a single $\sigma$) to accurately represent the varying connection strengths [e.g., \cite{Simpson-Porco2013}].
Also, the interaction strengths can evolve dynamically based on the states of the connected nodes through adaptive coupling laws \citep{dediga:09}.
Further complexity arises from intermittent or switching interactions, where connections between agents may be temporarily lost or deliberately modified, or instantaneously commutating between different values \citep{yang2016global,coraggio2018synchronization}.
These realistic features must be carefully considered when designing control strategies for practical applications.

%----------------------------
\subsection{Simplified Network Model}

While real systems exhibit considerable complexity, analysis often begins with simplifying assumptions. These typically include 
(i) considering identical agents ($f_i = f_j = f$, $\forall i,j$) ,
(ii) unitary response to interaction ($G_i = I$, $\forall i$), 
(iii) static pair-wise interactions,
and (iv) a linear diffusive coupling protocol ($h(x_i, x_j) = x_j - x_i$).
Under these assumptions, the network dynamics can be written compactly as 
\begin{equation}\label{eq:simplified_network}
    \dot{x}_i(t) = f(x_i(t)) - \sigma\sum_{j=1}^N L_{ij}x_j(t),
    \quad i \in \{1,\ldots,N\},
\end{equation}
where $L_{ij}$ are the elements of the graph Laplacian matrix.
This simplified model, while not capturing all the complexities of real systems, provides a tractable starting point for theoretical analysis and has proven remarkably useful in understanding fundamental properties of networked systems \citep{bullo2018lectures}.

\begin{figure}[t]
    \centering
    \includegraphics[width=1\linewidth]{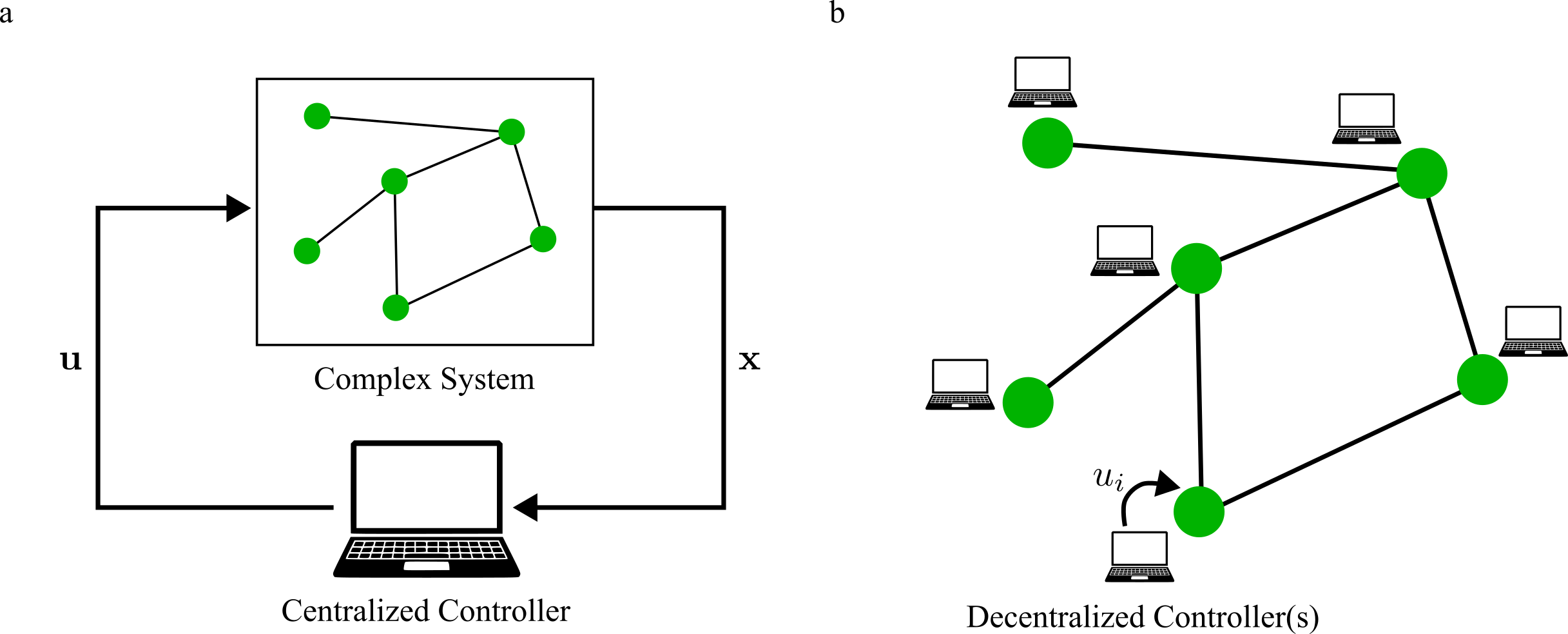}
    \caption{\textbf{Centralized and decentralized control strategies} 
    \textbf{a.} Schematic description of a centralized control strategy.
    In this paradigm, the controller is able to sense the entire state of the network $\B{x} \coloneqq [x_1\T \ \cdots \ x_i\T \ \cdots \ x_N\T]\T$ and potentially provide control inputs to every node through $\B{u} \coloneqq [u_1\T \ \cdots \ u_i\T \ \cdots \ u_N\T]\T$.
    \textbf{b.} Schematic description of a decentralized(/distributed) control strategy. Here, each controller measures only local information and provides a corresponding control input.}
    \label{fig:Centralized_decentralized}
\end{figure}

%---------------------------
\section{Controlling Complex Systems}

%-----------------------------
\subsection{Control Goals}

When controlling a complex system, we typically wish to enforce \emph{convergence} \citep{pavlov2004convergent} of the agents' states to a specific collective behavior.
One of the simplest ones is \emph{consensus}, where all agents converge to the same constant state \citep{OlFa:07}; 
this can be desired in rendez-vous problems of mobile agents, sensor fusion problems, and can occur in opinion dynamics networks \citep{amirkhani2022consensus}.
\emph{Synchronization} is arguably the most studied collective behavior and possibly the first being investigated, with Christiaan Huygens writing in 1673 about an “odd kind of sympathy” when he observed in the behavior of coupled pendulum clocks.
A synchronized complex system is one where all agents' states converge to the same solution, which can also be a non-constant function of time \citep{piro:01,arenas2008synchronization,boccaletti2018synchronization}.
In power grids, synchronization of the derivatives of power angles associated to generators is a crucial requirement for correct operation \citep{dorfler2014synchronization}.
In the brain, moderate levels of neural synchronization can be desired, as it is linked to motor coordination, whereas excessive levels of synchronization are related to seizures \citep{popovych2014control}.
To fit application needs, numerous particularizations of synchronization exist, including synchronization behavior that is \emph{bounded} (or \emph{practical}) \citep{vasca2021practical},
\emph{partial} (or \emph{output}) \citep{chopra2012output},
\emph{clusterized} \citep{lu2010cluster},
and of the type observed in \emph{chimera states} \citep{abrams2004chimera}

When agents' states have meaning akin to positions, velocities, and orientation, control goals related to spacial positioning are common. 
For instance, \emph{aggregation} (or \emph{cohesion}) tasks require agents to move as close as possible to one another \citep{gazi2003stability}, and when they move with similar velocity vector, they are said to \emph{flock} \citep{olfati-saber2006flocking}.
In \emph{formation control}, each agent must achieve a unique position relative to others \citep{oh2015survey,bayindir2016review}, whereas \emph{geometric pattern formation} aims to position the agents according to a structured arrangement, such as a tessellation of space \citep{giusti2023local}.
Finally, more complex behaviors find application in domains such as search and rescue or crowd control.
Examples include \emph{foraging} and \emph{area coverage} \citep{queralta2020collaborative}, or \emph{herding}, where only a subset of units can be controlled, and the goal is to influence the movement of the rest of the agents towards a desired one \citep{long2020comprehensive}.

In \emph{containment} or \emph{confinement} problems, typical of epidemics or flow network applications, the agents' states must remain confined in certain regions of the network's state space \citep{dellarossa2020network}; similar problems are sometimes framed as \emph{safety} problems in robotics \citep{wang2017safety}.
Finally, many decision-making problems over networks systems involve some form of optimization (e.g., power dispatch in power grids, traffic light control in road networks), although the agents are not always assumed to have a dynamics; we refer to \cite{thai2012handbook,nedic2018distributed} for more detailed discussions.

\begin{figure}[t]
    \centering
    \includegraphics[width=0.8\linewidth]{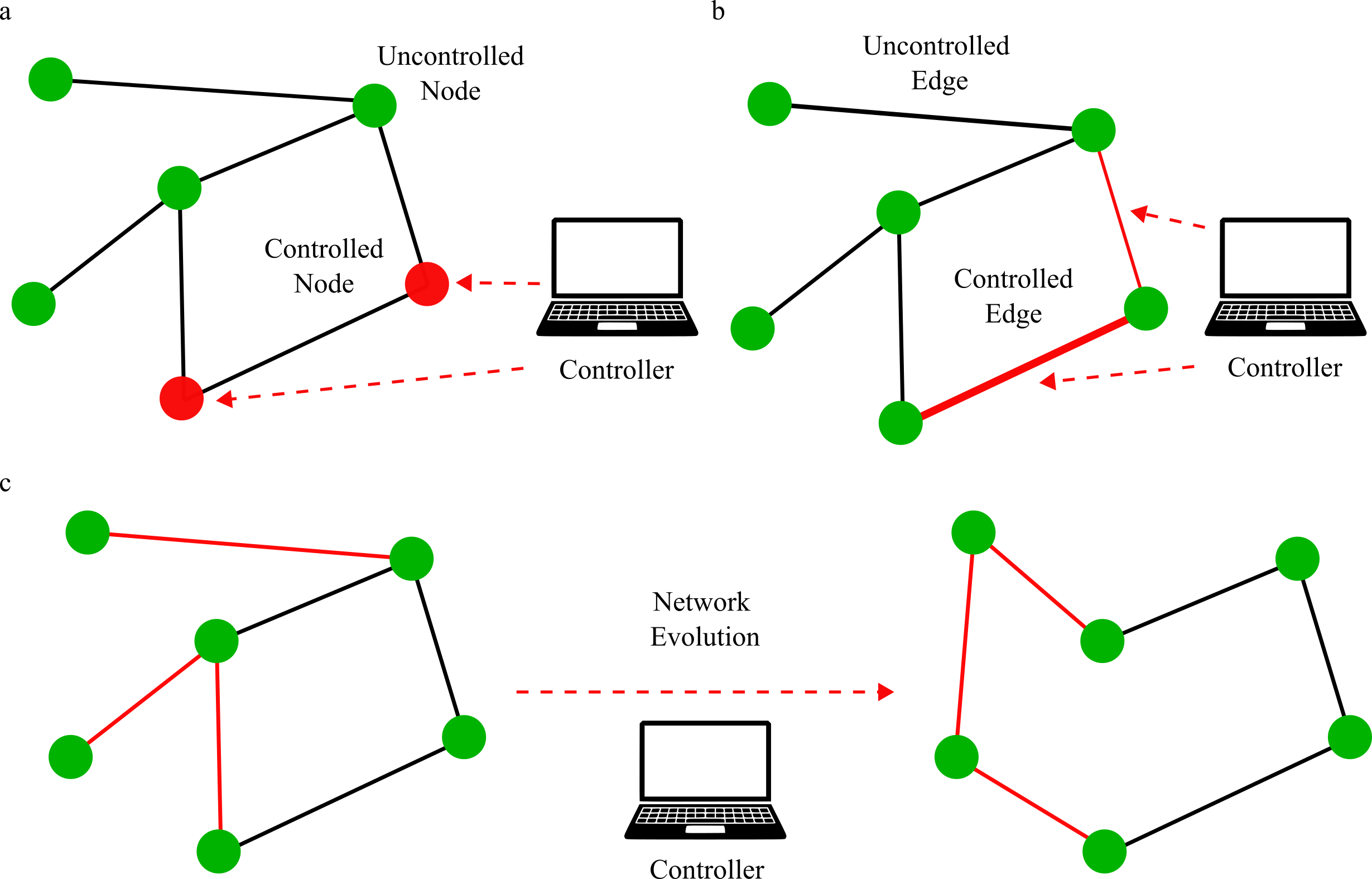}
    \caption{\textbf{Main control architectures} \textbf{a.} Schematic description of a node control strategy.
    Here, the controller directly influences the dynamics of a subset of the nodes of the network. \textbf{b.} Schematic description of an edge control strategy.
    This paradigm revolves around modifying the interaction protocol or coupling strength of a fraction of the edges of the network.
    \textbf{c.} Schematic description of a structural control strategy.
    Here, the controllers orchestrate the collective behavior of the complex system by dynamically modifying the structure of the interaction network.}
    \label{fig:Control_families}
\end{figure}

%------------------------------------------
\subsection{Control Approaches for Complex Systems}

The control of complex systems introduces distinct challenges that demand approaches fundamentally different from classical control theory. Instead of regulating a single system, we must shape the collective behavior of numerous interconnected agents through strategic interventions. This presents an inherently multiscale control problem: localized actions at the microscopic level of individual agents must propagate through the network to achieve desired macroscopic outcomes in the collective behavior.

A fundamental classification in network control strategies distinguishes between centralized and decentralized/distributed approaches. \emph{Centralized} control strategies, illustrated in Figure \ref{fig:Centralized_decentralized}a, rely on a single computing entity that determines control actions across the entire network, even when multiple actuation points are distributed among nodes. In contrast, \emph{decentralized}/\emph{distributed} control strategies employ multiple decision-makers operating with limited or local information, as shown in Figure \ref{fig:Centralized_decentralized}b. While this approach introduces additional constraints in control design, it offers enhanced robustness and reliability.%
\footnote{Different authors may propose more subtle differences between distributed and decentralized control strategies, also depending on the domain considered.}

\begin{figure}[t]
    \centering
    \includegraphics[width=1\linewidth]{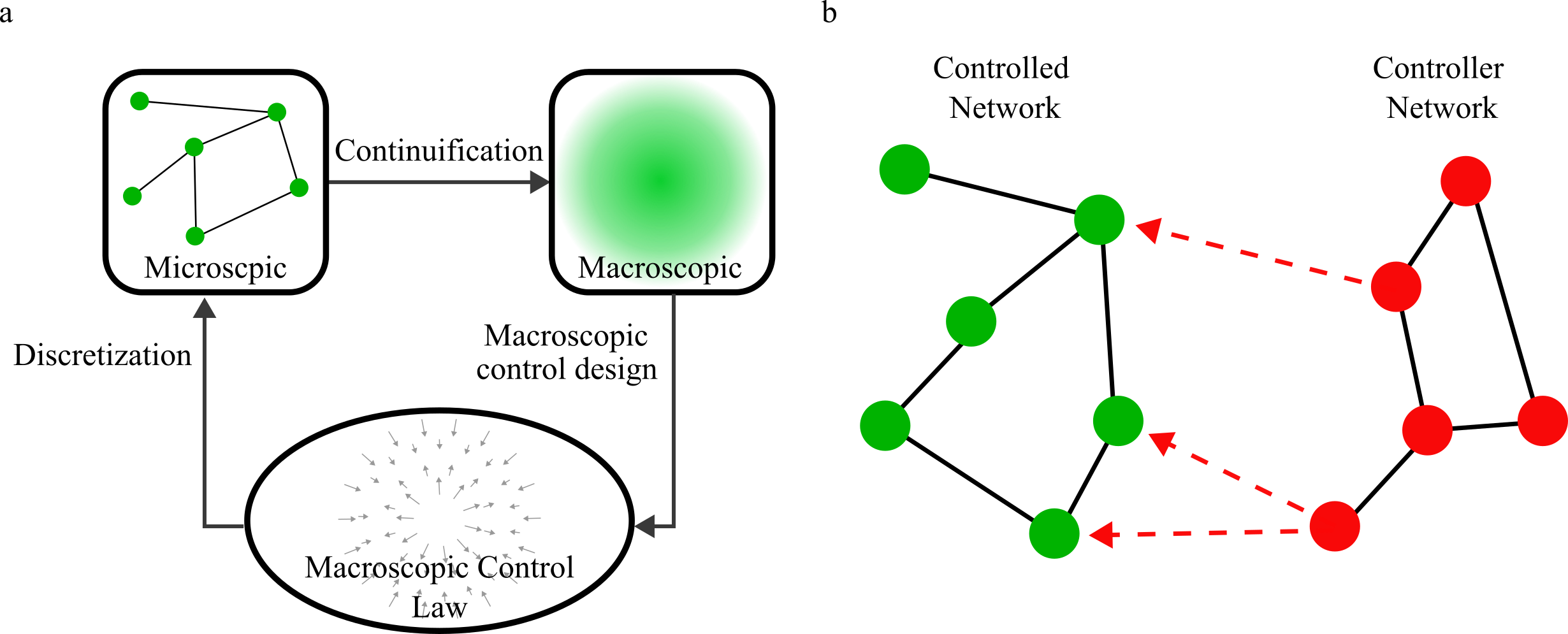}
    \caption{\textbf{Advanced control paradigms.} \textbf{a.} Schematic representation of a continuification based architecture to control a large-scale complex system.
    In this strategy a continuum model of the complex system is used to devise a macroscopic control law that is then discretized and applied to the agents of the network.
    \textbf{b.} Graphical description of a strategy that harnesses complex systems for control. 
    Here, a network of \emph{controller} agents, through local interactions, shapes the emergent behavior exhibited by a second network of \textit{controlled} agents.}
    \label{fig:Advanced_control_paradigms}
\end{figure}

Additionally, three distinct approaches characterize the control of complex systems: \emph{node control} (Figure \ref{fig:Control_families}a), which directly influences the dynamics of selected agents at the microscopic level; \emph{edge control} (Figure \ref{fig:Control_families}b), which modifies the interaction protocols between agents; and \emph{structural control} (Figure \ref{fig:Control_families}c), which dynamically alters the network topology to shape macroscopic behavior \citep{d2023controlling}.
While each approach has its merits, combining these strategies can often yield more effective solutions by operating simultaneously across multiple scales \citep{dedigo:10}. 
Indeed, modern applications, e.g., large-scale complex systems of mobile agents, are driving the development of new frameworks for controlling complex systems, such as that of continuification (see Figure \ref{fig:Advanced_control_paradigms}a) \citep{nikitin2021continuation,maffettone2023continuification}. 
Similarly, scenarios in which a group of controlled agents must cooperatively influence the behavior of a separate group---increasingly more common in applications like crowd and traffic management, or the confinement of pollutants---inspired the idea of harnessing complex systems for control (see Figure \ref{fig:Advanced_control_paradigms}b), as we will discuss below.

%------------------------------------------------
\subsection{Pinning Control: Learning from Nature}

Nature demonstrates how complex systems can achieve efficient control through localized interventions. Two examples illustrate this principle: in cardiac tissue, a relatively small population of pacemaker cells orchestrates the synchronized beating of the entire heart through local signal propagation; in honeybee swarms, a minority of scout bees guides the entire colony to new nesting sites through distributed communication protocols \citep{chen2014f}.
These natural examples inspired the development of \emph{pinning control} \citep{wang2002pinning}, where macroscopic behavior is influenced by controlling a small fraction of nodes. 
The key insight is that network interconnections can propagate the influence of controlled nodes throughout the system, making it possible to achieve global coordination through local interventions on a limited number of agents.

This approach can be mathematically described as follows. Consider a network of identical nonlinear systems described by equation \eqref{eq:simplified_network}. 
The pinning control strategy introduces one or more additional nodes, the \emph{leaders} or \emph{pinner nodes}, dividing the network into two communities: nodes directly controlled by the leaders (say $M$ nodes) and uncontrolled nodes (numbering $N-M$). Typically, $M\ll N$ (control can often be achieved with just 10--20\% of nodes being influenced by the pinner nodes).
The controlled nodes evolve according to
\begin{equation}
    \dot x_i(t) = f(x_i(t)) - \sigma \sum_{j=1}^{N} L_{ij}x_j(t) + u_i(t),
    \quad i \in \{1, \ldots, M\},
    \label{eq:Pinned Nodes}
\end{equation}
where $u_i : \mathbb{R} \to \mathbb{R}^n$ denotes the control action exerted by the pinner(s).
Meanwhile, the uncontrolled nodes follow their original dynamics:
\begin{equation}
    \dot{x}_i(t) = f(x_i(t)) - \sigma \sum_{j=1}^{N} L_{ij}x_j(t), \quad i \in \{M+1,\dots,N\}.
\end{equation}
A simple yet often effective control law is state feedback, where $u_i(t) = \sigma k_\mathrm{p}(x_\mathrm{p}(t)-x_i(t))$, with $x_\mathrm{p}(t)$ denoting the desired trajectory and $k_\mathrm{p}$ the control gain.
However, some applications may require more sophisticated control laws. 
Proposed strategies include adaptive pinning control \citep{guo2021event,turci2014adaptive}, optimal control \citep{della2023nonlinear}, approaches based on reinforcement learning \citep{rahnama2019learning} and more, generally inspired from nonlinear control strategies.

%---------------------------------------------
\subsection{Controllability and Observability of Complex Systems}

A fundamental question emerging from the framework of pinning control is: how many and which nodes should be controlled to ensure the control action steers the system towards the desired behavior? This problem, highlighted in the seminal work of \cite{LiSl:11} for the case of linear agent dynamics, has revealed surprising insights.
Contrary to intuition, only a small fraction of nodes typically needs to be controlled, and control often avoids hubs, favoring nodes with fewer connections, while the network structure significantly influences control effectiveness.

These counterintuitive findings highlight the importance of analytical frameworks to understand and predict under which conditions it is possible to control a network towards the desired trajectory. To address this challenge, researchers have developed two main approaches for analyzing network controllability. 
The first employs graph-theoretic methods that focus on identifying structural properties that guarantee controllability \citep{Mesbahi:2010aa,LiSl:11,yuan2013exact}, 
with extensions that are suitable for undirected and weighted graphs \citep{yuan2013exact}, provide less conservative results on the number of agents to control \citep{gao2014target}, provide information on the amount of energy required to solve a control problem \citep{yan2012controlling}, and balance the number of nodes to be controlled with the required control energy \citep{pasqualetti2014controllability}. 
The second approach is based on the master stability function, which analyzes the conditions for the local stability of desired states, as discussed for instance in \cite{sorrentino2007controllability}.
The dual problem of controllability, that is \emph{observability}, has also been tackled for complex systems, showing that it is indeed possible to observe the state of a complex system by measuring just part of the agents' states \citep{liu2013observability}.

%-------------------------------
\subsection{Advanced Node Control Strategies}

Beyond basic pinning control, several more sophisticated approaches have emerged to address the complexities of real-world networks.
Adaptive control strategies automatically adjust control gains to respond to changing conditions \citep{das2010distributed}, while robust control methods handle uncertainties and disturbances inherent in complex systems \citep{modares2019resilient}. 
For networks with time-varying topologies, switched or impulsive control approaches provide frameworks for maintaining performance despite structural changes, and hybrid control strategies combine continuous and discrete dynamics to handle delays and mixed-mode behaviors \citep{zhao2009synchronization,yang2011stochastic,zhou2012}.
Recent research has emphasized crucial practical considerations, including the impact of nonlinear node dynamics \citep{Jiang:2019bz} and the development of frameworks for scenarios that involve constrained node selection \citep{delellis2018partial}. 
Modern control architectures have become increasingly sophisticated, solving network optimization problems online \citep{coraggio2022minimax}, or incorporating multilayer and multiplex control strategies, enhancing performance through the interaction of multiple control layers with different topologies \citep{lombana2016multiplex}.
Other advanced methods span synchronization control \citep{scardovi2010synchronization}, cooperative control \citep{seyboth2016cooperative}, and distributed control approaches \citep{liuzza2016distributed}, with particular attention to practical stability \citep{montenbruck2015practical} and finite-time control objectives \citep{yang2015finite}.

Additionally, recent node control techniques are increasingly leveraging data-driven methods, employing system identification for network dynamics, machine learning for control policy design, and adaptive learning-based strategies \citep{baggio2021datadriven,asikis2022neural,d2023controlling,wu2024predicting}.
A large and growing area of study is that of the application of \emph{multi-agent reinforcement learning}, which applies principles of reinforcement learning (most often, temporal-difference approaches to solve optimal control problems) to complex networks, modeled as \emph{stochastic games}.
These approaches promise to solve very complicated control problem via machine learning, and have had remarkable success in teaching agents how to play competitive games \citep{vinyals2019grandmaster,openai2019dota}, how to play hide-and-seek in teams \citep{baker2020emergent}, and how to practice boxing \citep{won2021control}.
However, strong theoretical (and often computational) limitations persist, induced by nonstationarity of the agents' control policies, partial observability, and credit assignment. 
We refer to \cite{busoniu2008comprehensive,nguyen2020deep} for more a detailed description.

%-----------------------------------------
\subsection{Edge Control}

Real-world complex systems rarely conform to simplified assumptions of identical agents, uniform coupling strengths, and static topologies. 
Operating in environments characterized by noise, disturbances, and time-varying conditions, these systems have driven the development of increasingly sophisticated modeling and control frameworks. Specifically, the development of frameworks to describe time varying networks has put the basis for the development of strategies that aim to coordinate the behavior of the system by modifying the interaction protocols and/or the structure of the network.
Among the landmark developments that allowed the theoretical understanding of dynamic networks, there is Siljak's pioneering work in the 1970s \citep{Si:78,Siljak:2008p3179}, which first formalized systems with dynamical nodes and edges, though maintaining fixed structure \citep{Zecevic:2010p1970}. 

A crucial advancement has been the development of adaptive coupling strengths. 
While maintaining a fixed network structure, edges become dynamical elements whose strengths evolve based on the states of connected nodes:
\begin{equation}
    \dot{x}_i(t) = f_i(x_i(t)) - \sum_{j=1}^N \sigma_{ij}(t) L_{ij} x_j(t),
    \quad i \in \{1,\ldots,N\},
\end{equation}
where time-varying coupling strengths $\sigma_{ij}(t) \in \mathbb{R}$ follow appropriate adaptation laws \citep{dediga:09,lulu:06,YuDe:12,das2010distributed,kim2017adaptation}. This adaptive approach enables automatic tuning of coupling strengths, compensates for heterogeneities, enhances synchronization properties, and improves robustness to disturbances.
Additionally, it has been demonstrated that adaptive strategies can bridge one of the crucial gaps between microscopic and macroscopic control. 
As shown in \cite{kempton2017distributed,kempton2017self}, distributed estimation and adaptation mechanisms can control macroscopic network properties through targeted actions on individual edges.

To compensate for heterogeneity and discontinuity in agent dynamics, discontinuous and set-valued coupling have effectively been used to enforce synchronization \citep{coraggio2021convergence,coraggio2020distributed,miranda2024robust}.
Integral coupling protocols have also proved effective in the control of heterogeneous agents \citep{lombana2015}.
Finally, impulsive coupling protocols have been used synchronize agents subject to delays
\citep{yang2016global}.

%-----------------------------------------
\subsection{Structural Control}

Holland's introduction of complex adaptive systems \citep{HHolland:1975p3207} expanded the vision of adaptive edge weights to encompass systems that modify both their structure and their dynamics in response to environmental changes. 
More recent frameworks of adaptive networks \citep{Gross:2009p5768} and evolving dynamical networks \citep{Gorochowski:2010p5044} provide comprehensive treatments of systems with dynamic topologies and adaptive behaviors.
Edge-snapping control \citep{dedigapo:09} represents a sophisticated approach, enabling networks to dynamically rewire themselves. 
Edge dynamics follows that of a nonlinear bistable system:
\begin{equation}
\ddot{\sigma}_{ij}(t) + d\dot{\sigma}_{ij}(t) + \frac{\mathrm{d}V}{\mathrm{d}\sigma_{ij}}(\sigma_{ij}(t)) = g(x_i(t),x_j(t)),
\end{equation}
where $V(\sigma_{ij}(t))$ is a bistable potential function, $g(x_i(t),x_j(t))$ measures node state mismatch, and $d$ controls transient behavior. This framework enables dynamic creation and removal of connections, self-organization of network structure, and adaptation to changing conditions.

Implementation of these advanced strategies faces several practical challenges. 
These include managing computational complexity of network evolution, meeting communication requirements for distributed adaptation, analyzing stability of time-varying structures \citep{zhao2009synchronization}, ensuring robustness to noise and disturbances, and maintaining scalability to large networks.
The field continues to evolve rapidly, with new theoretical frameworks and practical applications emerging regularly; further notable examples of networks with time-varying structure include switched systems, networks with proximity graphs, and activity-driven networks \citep{ghosh2022synchronized}.

A complementary perspective to structure control is \emph{structure design}, where rather than changing the structure online, an optimal one is designed a priori, which can be the only option available for costly infrastructure such as road networks or power grids.
As a matter of fact, the propensity of a complex system to display a certain collective behavior is heavily influenced by its structure, as captured by metrics such as the
\emph{eigenratio}, the \emph{algebraic connectivity} and the \emph{minimum density}
\citep{donetti2005entangled,dedi:11,coraggio2021convergence}.
Both model-based \citep{fazlyab2017optimal} and data-based \citep{coraggio2024datadriven} approaches exist to carry out optimal structure design; however, the main challenge stems from the fact that different control goals can require completely different structure solutions and even design algorithms.

%----------------------------------------------
\subsection{Control of Large-Scale Systems}

In many real-world applications, the number of agents in the network can become very large%
\footnote{Which order or magnitude can be considered ``large'' depends on the application.}
(e.g., crowds, traffic or biological networks), with thousands of interacting agents. 
As the network size increases, the microscopic description of agent dynamics and interactions becomes both analytically and numerically challenging. 
In recent years, a technique known as \emph{continuification} has been developed as a promising solution to this problem \citep{nikitin2021continuation}.
The basic idea is to transform microscopic, agent-level dynamics into a macroscopic continuum description using partial differential equations (PDEs), design the control action at the macroscopic level, and then discretize it for application to the microscopic agents.
This approach facilitates the synthesis of control actions aimed at achieving desired global configurations while implementing controls at the local, individual-agent level.
Continuification methods are particularly useful in systems with large numbers of agents because they enable effective control over emergent collective behavior, thereby reducing the overall complexity of the control problem.
Applications include swarm robotics \citep{maffettone2023continuification, rubenstein2014programmable,gardi2022microrobot}, traffic systems \citep{siri2021freeway}, biological cell populations \citep{de2022control}, and crowd dynamics \citep{albi2020crowd}.

Several contemporary approaches use macroscopic representations of multi-agent dynamics. 
\emph{Mean-field} control techniques \citep{elamvazhuthi2021controllability, borzi2020mean} approximate the behavior of large agent ensembles by focusing on the mean interaction across the system, while \emph{graphon}-based methods \citep{jordan2013mean} use continuum limits of large networks to enable scalable control strategies. 

The continuification approach can be formalized as follows.
Consider a system of \(N\) agents, each characterized by a position \(x_i(t) \in \Omega \) and a control input \(u_i(t) \in \mathbb{R}\) in a periodic domain \(\Omega\). The dynamics of the \(i\)-th agent can be expressed as
\begin{equation}
    \dot{x}_i(t) = \sum_{j=1}^{N} h(x_i(t), x_j(t)) + u_i(t),
\end{equation}
where here \(h : \Omega \times \Omega \to \mathbb{R}^n\)  represents the \emph{interaction kernel} modeling pairwise interactions, and $u_i \in \mathbb{R}^n$  is the control input. 
Assuming a sufficiently large \(N\), we describe the macroscopic behavior of the system with a density function $\rho: \Omega \to \mathbb{R}$ representing the agent distribution over \(\Omega\), satisfying the conservation law
\begin{equation}
    \frac{\partial \rho(z,t)}{\partial t} + \nabla \cdot \left[\rho(z,t) (V(z,t) + U(z,t)) \right] = 0,
\end{equation}
where $z$ represents the absolute position in $\Omega$, 
\(V(z, t) = \int_\Omega h(z, y) \rho(y, t) \, \, \mathrm{d}y\) is the velocity field generated by the interactions \citep{maffettone2023continuification},
and $U(z,t)$ is the velocity field generated by the control inputs $u_i$.
Using such a macroscopic description, it is possible to devise control laws that drive the density of the population to a desired configuration with provable convergence guarantees, such as the ones developed in \cite{maffettone2023continuification,maffettone2023mixed}. Additionally, the description of the network as a set of partial differential equations paves the way for the extension of tools coming from the PDE control control literature \citep{krstic2008boundary} for the control of large-scale complex systems.

%---------------------------------------------
\subsection{Harnessing Complex Systems for Control}

All the solutions presented until now focus on designing external controllers to influence networked systems. An alternative and powerful approach involves harnessing one complex system to control another. 
In this framework, a group of controller agents is responsible for controlling another population of uncontrolled target agents. 
A paradigmatic example of this is the \emph{shepherding} problem, where a group of agents (\emph{herders}) must coordinate to control the collective dynamics of another group (\emph{targets}). 
This scenario appears naturally in biological systems, such as dolphins hunting fish \citep{haque2009hybrid} or ants collecting aphids, and has important applications in technological domains including search and rescue operations, crowd control, and environmental cleanup.

In its essence, this control framework revolves around designing $M$ controllers tasked with coordinating the behavior of $N$ targets. 
In general, denoting with $x_{\mathrm{t},i}$ the state of the $i$-th target, this population can be microscopically modeled as
\begin{equation}
    \dot{x}_{\mathrm{t},i} = f_\mathrm{t}(x_{\mathrm{t},i}) + \sum_{j=1}^N h_{\mathrm{t}\mathrm{t}}(x_{\mathrm{t},i},x_{\mathrm{t},j}) + \sum_{j=1}^M h_{\mathrm{t}\mathrm{c}}(x_{\mathrm{t},i},x_{\mathrm{c},j})
\end{equation}
where $h_{\mathrm{t}\mathrm{t}}$ and $h_{\mathrm{t}\mathrm{c}}$ are the interaction protocols between controllers and targets, and $f_\mathrm{t}$ describes the individual dynamics of the targets.
Similarly, the controllers' dynamics is described by
\begin{equation}
    \dot{x}_{\mathrm{c},i} = f_\mathrm{c}(x_{\mathrm{c},i}) + \sum_{j=1}^M h_{\mathrm{c}\mathrm{c}}(x_{\mathrm{c},i},x_{\mathrm{c},j}) + \sum_{j=1}^N h_{\mathrm{t}\mathrm{c}}(x_{\mathrm{t},j},x_{\mathrm{c},i}) + u_i
\end{equation}
where $f_\mathrm{c}$ represents the controllers individual dynamics, $h_{\mathrm{c}\mathrm{c}}$ is the interaction protocol between the controllers, and $u_i$ describes the control input.
Different control solutions have been proposed for this control problem, ranging from heuristic rules \citep{lama2024shepherding,nalepka2019human} to optimal control solutions \citep{escobedo2016optimal}.
This framework has several important implications for the design of multi-agent control systems. 
It provides guidelines for determining the minimum number of control agents needed \citep{lama2024shepherding}, clarifies the relationship between local interactions and global control objectives, and offers insights into how system performance scales with the number of agents \citep{maffettone2024leader}. 
These findings are particularly relevant for large-scale distributed systems where centralized control is impractical or impossible.

Several open challenges remain in harnessing complex systems for control.
These include developing continuum models to describe emergent shepherding behavior, engineering local interaction rules for more complex tasks, addressing scenarios with actively escaping targets, and extending the framework to three-dimensional spaces and other geometries.

%-------------------------------
\section{Proving Stability}

The paramount property of a feedback controller is its ability to ensure that the controlled system converges to desired behavior and maintains it despite disturbances.
Formally, this normally means that a specific region in the state space of the network is made invariant and asymptotically stable \citep{meiss2017differential}.
Thus, it is fundamental to establish in which conditions controlled complex networks are stable. 
Multiple theoretical frameworks have emerged to address this challenge, each offering distinct insights and capabilities while complementing the others.

%-------------------------------------------------
\subsection{Global Stability through Lyapunov Theory}

Lyapunov-based methods offer powerful analytical tools to prove local or global stability of invariant solutions \citep{dedi:11}.
The approach centers on constructing suitable candidate Lyapunov functions \citep{khalil2002nonlinear}.
These are scalar functions $V:\mathbb{R}^{Nn}\to\mathbb{R}$ that are continuous and differentiable, and positive everywhere but in the desired state $x^*$.
Then, to ensure that the trajectories of the system converge to the desired one, it is necessary to establish conditions so that
\begin{equation}
\dot{V}(x) < 0, \quad \forall x \neq x^*.
\end{equation}
Numerous extensions to this theory exist, to prove convergence to invariant sets that are not points, with non-autonomous systems, etc. \citep{khalil2002nonlinear}; when inputs must directly be taken into account, \emph{passivity} theory can be employed \citep{arcak2007passivity}.
The main difficulty associated with Lyapunov based methods is finding a suitable function $V$.
While these methods can handle nonlinear vector fields and provide global results, they often require additional conditions to impose some regularity on the dynamics, such as the \emph{one-sided Lipschitz} (or \emph{QUAD}) condition (and its extensions) \citep{bullo2022contraction,coraggio2020analysis}, and the resulting stability criteria can be conservative.
Nevertheless, their analytical nature makes them particularly valuable for theoretical analysis.

%------------------------------------
\subsection{Dynamical Analysis through Contraction Theory}

Contraction theory \citep{losl:98,slwa:98,bullo2022contraction} offers a powerful alternative focusing on the convergence of trajectories through analysis of the differential dynamics
\begin{equation}
\dot{\delta x}(t) = J(x,t) \delta x(t),
\end{equation}
where $\delta x(t)$ is a virtual displacement with respect to $x^*(t)$, that is the desired trajectory, and $J \coloneqq \tfrac{\partial f}{\partial x}(x(t),t)$ is the Jacobian matrix of the system described by $\dot x(t) = f(x(t),t)$. 
The theory shows that if there exists a \emph{matrix measure} (also known as \emph{logarithmic norm}) \citep{soderlind2006logarithmic} of $J$ that is negative, then all trajectories converge toward each other (\emph{incremental stability}).
This approach proves particularly effective for certain system classes, especially biological networks \citep{rusl:10,rudi:10}, offering several advantages.
First, it provides a natural framework for synchronization analysis, directly addressing the convergence of trajectories rather than stability of particular solutions (which need not be known a priori) \citep{russo2013contraction}.
Second, it allows the use of different matrix measures, providing flexibility in analyzing various types of systems.
Finally, it readily handles time-varying dynamics, making it suitable for adaptive and evolving networks.

%-----------------------------------------------
\subsection{Local Stability in Synchronization Problems: The Master Stability Function}

When the desired emergent behavior is the synchronization of all the agents onto the same trajectory, the \emph{master stability function} (MSF) approach, pioneered by \cite{peca:98} [see also \cite{fujisaka1983stability}], provides a powerful framework for analyzing local stability.
The method proceeds through systematic reduction of the high-dimensional stability problem to a parametric analysis of individual modes. 
Namely, linearization of equations \eqref{eq:general_network}--\eqref{eq:coupling} about the synchronous solution $x_{\mathrm{s}}(t)$ is considered:%
\footnote{We assumed $f_i = f$ and $G_i = I$ for all $i$, and $h(x_i, x_j)$ writeable, with slight abuse of notation, as $h(x_j) - h(x_i)$; moreover, $\sum_{j=1}^N A_{ij}\,(h(x_j) - h(x_i)) = -\sum_{j=1}^N L_{ij} h(x_j)$.}
\begin{equation}
\dot{\xi}_i(t) = \frac{\partial f}{\partial x}(x_{\mathrm{s}}(t))\ \xi_i(t) - \sigma\sum_{j=1}^N L_{ij} \frac{\partial h}{\partial x}(x_{\mathrm{s}}(t))\ \xi_j,
\end{equation} 
where $\xi_i(t) = x_i(t) - x_{\mathrm{s}}(t)$ is the synchronization error. 
The approach employs then block diagonalization; specifically, one lets $\B{\zeta} \coloneqq (Q^{-1} \otimes I) \B{\xi}$, where $Q$ is the modal matrix of $L$, $\otimes$ is the Kronecker product, and $\B{\xi}, \B{\zeta} \in \mathbb{R}^{Nn}$ are stack vectors.
Thus, it is possible to obtain the variational equation
\begin{equation}\label{eq:diagonalized_msf}
\dot{\zeta_i} = \left[ \frac{\partial f}{\partial x}(x_\mathrm{s}) - \alpha \frac{\partial h}{\partial x}(x_\mathrm{s}) \right] \zeta_i,
\end{equation}
where $\alpha \coloneqq \sigma\lambda_i$; $\lambda_i$ being the $i$-th eigenvalue of the Laplacian matrix. 
The MSF, typically denoted as $\Lambda(\alpha)$, is the largest Lyapunov exponent of Equation \eqref{eq:diagonalized_msf}.
Hence, the synchronization manifold is locally asymptotically stable if $\Lambda(\alpha) < 0$ for $\alpha \in \{\sigma \lambda_2, \dots, \sigma \lambda_N\}$.

This approach elegantly combines network structural properties (through the spectrum of the Laplacian matrix) with node dynamics, providing explicit conditions on coupling strength required for synchronization.
While semi-analytical in nature, requiring numerical computation of Lyapunov exponents, it enables systematic classification of synchronizability for different network topologies.
Recent extensions extended the use of the master stability function to study synchronization of piecewise-smooth systems \citep{dieci2023master} and the transitions from cluster to full synchronization
\citep{bayani2024transition}.

%------------------------------------------------
\subsection{Unified Understanding through Comparative Analysis}

These approaches complement each other in several key aspects.
In terms of scope, the MSF provides detailed local stability conditions, and tend to be less conservative and more flexible with respect to different interaction protocols, while Lyapunov methods and contraction theory can provide also global results, but tend to be more conservative, especially depending on which Lyapunov function or contraction norm are selected. 
Computationally, the approaches differ significantly: the MSF requires numerical computation of Lyapunov exponents, Lyapunov methods often yield analytical results but require appropriate selection of a valid Lyapunov function, and contraction theory enables a hybrid approach combining analytical insights with numerical verification. 
Each method also shows distinct strengths in different applications: the MSF excels for networks of identical nodes, Lyapunov methods handle heterogeneous systems well, and contraction theory proves particularly effective for certain dynamical structures common in biological and physical systems.

Recent developments have significantly expanded these foundational approaches.
Unified frameworks now combine different methodologies \citep{khong2016unifying}, while extensions address time-varying networks \citep{Wieland2011} and nonlinear systems \citep{andrieu2018some}.
New methods for analyzing partial and cluster synchronization have emerged, broadening the scope of application \citep{zhang2021unified,della2020symmetries}.
However, several fundamental challenges persist. 
The analysis of heterogeneous networks remains difficult, particularly when considering convergence in adaptive and evolving networks.
Understanding the impact of noise and uncertainties, developing scalable methods for large networks, and establishing tight bounds on convergence rates represent ongoing challenges. 
A comprehensive treatment of these developments and challenges can be found in \cite{di2016convergence}.

%--------------------------------------
\section{Summary and Future Directions}

Complex systems in modern applications are characterized by three fundamental components: node dynamics describing individual agent behavior, interaction functions governing inter-agent influences, and network structure defining interconnection topology. These systems exhibit remarkable emergent properties, with synchronization representing the most extensively studied collective behavior. Control strategies have evolved along three complementary paths: direct influence on selected agents through node control, modification of interaction dynamics via edge control, and dynamic rewiring of network topology through structural control. Recent advances have expanded these approaches to address large-scale systems and harness complex networks for control purposes.

Despite significant progress, multiple challenges persist.
Heterogeneity presents a fundamental obstacle, particularly in biological systems where network components vary significantly.
Time-varying structures in robotics and modern power grids demand novel theoretical frameworks.
Practical implementation must address noise, uncertainties, and communication complexities, including intermittent connections, limited range, and packet drops.
The theoretical framework must expand to encompass heterogeneous stochastic systems, strongly nonlinear or hybrid agents, and the role of noise in control \citep{Rus_Sho_16a,Burbano2017}.

The field is advancing toward sophisticated paradigms emphasizing self-organization through adaptive network structures, interacting complex systems, and distributed decision-making. Bio-inspired strategies derive crucial insights from natural systems—from fish schools performing coordinated maneuvers to bird flocks maintaining formations and social insect colonies exhibiting collective intelligence. Advanced control architectures increasingly incorporate hybrid strategies, multi-layer structures, and adaptive, learning-based approaches. Furthermore, compact descriptions of emergent behavior in large-scale networks have established foundations for applying PDE control to complex systems.

Applications span diverse emerging fields. In biology, these range from cell colony coordination to tissue engineering and synthetic biology. Robotics applications encompass swarm systems, microrobot collectives, and autonomous vehicle formations. Infrastructure benefits through applications in smart power grids, transportation networks, and communication systems.

Future research focuses on developing control strategies with enhanced capabilities: self-organization and adaptation to changing conditions, robust operation in uncertain environments, efficient scaling to large networks, effective function with limited communication, and achievement of complex collective behaviors. This represents a fundamental shift in complex system management.
This field continues to expand control theory's frontiers, offering opportunities for both fundamental research and practical applications. Success will enable new technologies in swarm robotics, biological engineering, and fields where coordinated collective behavior is essential. As we uncover principles governing complex collective behavior, we approach control systems matching the sophistication of natural systems, opening possibilities from medicine to robotics where numerous units must coordinate to achieve common objectives.

\bibliography{Biblio}

\end{document}